\documentclass{aa}
\usepackage[varg]{txfonts}
\usepackage{siunitx}
\usepackage{graphicx}
\usepackage{hyperref}

\bibpunct{(}{)}{;}{a}{}{,}

\hypersetup{ 
    colorlinks,
    linkcolor=blue,
    citecolor=blue
}

\newcommand{\dv}[2]{\frac{\textrm{d} #1}{\textrm{d} #2}}

\newcommand{\iunits}{\unit{kW.m^{-2}.nm^{-1}.sr^{-1}}}

\begin{document}

\title{Irregular grids for 3D NLTE radiative transfer in stellar atmospheres}

\author{Elias R. Udn{\ae}s
        \inst{1,2}
        \and Tiago M.D. Pereira
        \inst{1,2}
        }
\institute{Rosseland Centre for Solar Physics, University of Oslo, P.O. Box 1029 Blindern, NO--0315 Oslo, Norway
\and
Institute of Theoretical Astrophysics, University of Oslo, P.O. Box 1029 Blindern, NO--0315 Oslo, Norway}

\abstract 
{Three-dimensional non-local thermodynamical equilibrium (NLTE) radiative transfer calculations are a fundamental tool for a detailed spectral analysis in stellar atmospheres, but require vast amounts of computer power. This prevents their broader application.} 
{We undertake a first exploration of the use of 3D irregular grids in stellar atmospheres. In particular, we aim to test whether irregular grids can be used to speed up the 3D NLTE problem, in the same way as depth optimisation can lead to faster running times in 1D.}
{We created irregular grids based on 3D Voronoi diagrams, sampling different distributions from a 3D radiation-magnetohydrodynamic Bifrost simulation. We developed a method for solving radiation on the 3D irregular grid and implemented a simple NLTE solver using $\Lambda$-iteration and statistical equilibrium. We applied this to a simplified hydrogen-like atom and studied the convergence properties and accuracy of the irregular grid methods. For reference, we compared them to a standard short-characteristics solver on a regular grid.}
{We find that our method for radiation in irregular grids gives similar results to those from regular grids, and that it is possible to obtain nearly the same results with about ten times fewer points in the irregular grid for the continuum intensity in local thermodynamical equilibrium. We find that the irregular grid can give good results for the NLTE problem, but it takes four times longer per iteration than the regular grid, and it converges in about the same number of iterations. This makes it particularly inefficient. Our formulation therefore does not lead to an improvement. We also find that the design of the irregular grid is crucial for accurate results, and find it non-trivial to design an irregular grid that can work well across a wide range of heights.}
{}

\date{}

\keywords{Radiative transfer -- Methods: numerical -- Line: formation -- Sun: atmosphere -- Stars: atmospheres}

\maketitle

\section{Introduction}

Radiative transfer calculations are essential in stellar spectroscopy. They are the basic building block of spectral synthesis from model atmospheres, which are used to infer the temperature, chemical composition, magnetic fields, and many other quantities. In optically thick stellar atmospheres, detailed calculations outside local thermodynamical equilibrium (LTE), or non-LTE (NLTE), are typically solved iteratively \citep[e.g.][]{Cannon:1973un,Rybicki:1991ws}. To ensure convergence of these schemes, we need to accurately evaluate the radiation field at all grid points (known as a formal solution). Radiation is sensitive to changes in optical depth. Numerically, the formal solution is most stable when there are no large jumps in optical depth between the grid points. However, this is often not the case in model atmospheres, whose grids are optimised to solve the hydrostatic or magnetohydrodynamic equations, and not the transport of radiation.

To increase the accuracy in radiative transfer in stellar atmospheres, different strategies are used. The brute-force approach would be to interpolate the model to a much finer grid, but this raises many problems. First, it can dramatically increase the computational effort because we need to integrate over potentially many more points. Second, iterative NLTE schemes need to propagate a change in the solution, which is linearly proportional to the number of depth points \citep[see e.g.][]{Olson:1986ul, Stepan:2022aa}. Therefore, it is highly desirable to solve the formal solution without substantially increasing the number of grid points. One approach is to use higher-order interpolation \citep[e.g.][]{Auer:1976, Trujillo-Bueno:2003aa, Holzreuter:2012aa, de-la-Cruz-Rodriguez:2013aa, Janett:2019ab}, although these approaches still struggle in coarse grids or discontinuous media \citep{Steiner:2016aa}. Another approach, used in some 1D codes \citep[e.g. in the current version of the MULTI code;][]{Carlsson:1986} is to interpolate from the original grid to a grid that resolves the gradients in optical depth and other associated quantities better. Yet another approach is to use multi-grid iterations, where solving the radiation on a coarser grid is used to remove the high-frequency components of the error of fine grid solutions \citep[e.g.][]{Fabiani-Bendicho:1997vl, Stepan:2013aa, Bjorgen:2017}.

With time-dependent 3D model atmospheres being widely used, it becomes necessary to carry out radiative transfer calculations in 3D. Compared to the classical 1D model atmospheres, these calculations are computationally much more demanding and require efficient algorithms. Assuming the same number of grid points $N$ in the three spatial dimensions, together with a number of angles $N_\Omega$ and number of wavelengths $N_\lambda$, \citet{Stepan:2022aa} estimated that the solution time of a 3D NLTE problem is about $O(N^4 N_\Omega N_\lambda)$. This is a great challenge for current problems in solar physics, in particular, for spectral lines that form in the chromosphere. Here, NLTE calculations are essential, and so is spatial resolution, given that the chromospheric plasma is organised into much finer scales than in the photosphere. This leads to exploding computational costs, needing many iterations to obtain the converged radiation field, and with computational costs often nearing millions of CPU hours for a single simulation snapshot \citep{Sukhorukov:2017, Bjorgen:2017}.

The aim of this work is to explore a different approach to solving 3D NLTE problems in stellar atmospheres: irregular grids. In 1D, using a grid that is better suited to the radiative transfer problem can improve the convergence of NLTE problems \citep[see e.g.][]{RH15D}. However, this approach cannot be directly applied to the 3D case in regular grids because we would need an optimised grid for every ray. By relaxing the requirement of a regular grid and placing grid points freely in 3D space, we can optimise the grid to the radiative transfer problem. 

Irregular grids are not new in astrophysical 3D radiative transfer problems, in particular, for cosmological simulations and clumpy media \citep[e.g.][]{Ritzerveld:2006aa,Paardekooper:2010um}. The \textsc{SimpleX} algorithm of \citet{Ritzerveld:2006aa} was designed to solve continuum radiative transfer in irregular grids created from Voronoi tessellation \citep{Dirichlet:1850ub, Voronoi:1908no}. 
\citet{Brinch:2010wj} used a similar approach in the \textsc{Lime} code, but extended it to include line opacity and to solve NLTE problems. In these algorithms, rays travel only between Delaunay lines, which are the lines that connect the cell centres of Voronoi cells. \citet{Camps:2013uu} developed another approach in which rays are allowed to travel on a straight path through a 3D Voronoi grid. All these previous studies find that using irregular grids can give accurate results (or results that are comparable to existing methods) using fewer grid points. For their problem, \citet{Camps:2013uu} find that 3D radiative transfer through a Voronoi grid is about three times slower than on a regular grid with the same number of cells, but that fewer cells are required to reach an accurate result. This outweighs the slower calculation times. 
\citet{Bruls:1999aa} developed methods for solving radiative transfer on unstructured grids and applied them to estimate the radiative heating in stellar atmospheres, but their approach is limited to LTE and 2D geometry.
To the best of our knowledge, no study has employed irregular grids in 3D NLTE problems in optically thick stellar atmospheres. The motivation for our work is therefore to investigate whether methods like this could be of value in stellar atmospheres, in particular, to speed up the convergence of 3D NLTE schemes.

\section{Methods\label{sec:background}}

\subsection{NLTE radiative transfer}

We aim to solve the time-independent radiative transfer equation, which for a given ray can be written as
\begin{equation}
    \dv{I_\lambda}{\tau_\lambda} = S_\lambda - I_\lambda,
    \label{eq:RTE}
\end{equation}
where $S_\lambda$ is the source function, and $\tau_\lambda$ is the optical depth.
The formal solution for the intensity $I_\lambda$ can be written as
\begin{equation}
    I_\lambda (\tau_\lambda) = I_\lambda (0) e^{-\tau_\lambda} + \int_0^{\tau_\lambda} S_\lambda(\tau_\lambda') e^{-(\tau_\lambda - \tau_\lambda')}\,\textrm{d}\tau_\lambda'\,.
    \label{eq:Formal}
\end{equation}

In an NLTE problem, $S_\lambda$ will depend on $I_\lambda$ via the angle-averaged intensity $J_\lambda$. Using the two-level atom simplification, this can be expressed as
\begin{equation}
    S_\lambda(J_\lambda, T) = (1 - \varepsilon_\lambda)J_\lambda + \varepsilon_\lambda B_\lambda(T)\,,
    \label{eq:Source_scattering}
\end{equation}
where $\varepsilon_\lambda$ is the photon destruction probability, and $B_\lambda(T)$ is the Planck function. Numerically, the $\Lambda$ iteration procedure involves iteratively solving Eq. (\ref{eq:Source_scattering}) until $S_\lambda$ is consistent with $J_\lambda$. For a given estimate of $S_\lambda$, we need to solve Eq. (\ref{eq:Formal}) for multiple directions to obtain $J_\lambda$, which can be expressed by the $\Lambda$ operator,
\begin{equation}
    J_\lambda = \Lambda [S_\lambda].
\end{equation}
This basic scheme is not very efficient and struggles in cases with strong scattering ($\varepsilon_\lambda < 10^{-3}$). Accelerated schemes for $\Lambda$ iteration (ALI) have been developed that are much more efficient \citep[e.g.][]{Cannon:1973, Rybicki:1991ws}. However, given the exploratory nature of our analysis, we adopted the simple $\Lambda$ iteration because it is faster to implement, and we restricted our problem to cases in which $\varepsilon_\lambda > 10^{-3}$ by artificially modifying the model atom. 

Efficient solutions of the 3D NLTE problem can involve methods to speed up the formal solution and/or reduce the number of iterations needed to reach a converged solution. In this work, we address the latter.

\subsection{Model atom and extinction sources} \label{sec:two-level}

To test radiative transfer on irregular grids, we made use of a simple two-level plus continuum model atom to keep the computational complexity down. We built our simplified model atom based on the hydrogen atom, but with some important differences. First, we used the first two levels of hydrogen, so there is only one spectral line (Lyman-$\alpha$-like), and two bound-free transitions. Second, we artificially increased the collisional excitation and ionisation rates, originally derived from \citet{Johnson:1972aa}, by a factor of $10^9$, to decrease the amount of scattering so that we were able to reduce the number of $\Lambda$ iterations and use the simpler $\Lambda$-iteration scheme. In our model, the lowest value of the destruction probability was $\varepsilon_\lambda \approx 0.09$.

\begin{table}
    \centering
    \caption{Energy levels of the adopted model atom.}
    \begin{tabular}{r c} 
    \hline\hline
    $E$ & $g$ \\
    $(\unit{cm^{-1}})$  & \\ [0.5ex] 
    \hline
    $0.000$ & 2  \\ 
    $82\,258.211$ & 8  \\
    $109\,677.617$ & 1\tablefootmark{a}\\
    \hline
    \end{tabular}
    \tablefoot{\tablefoottext{a}{\ion{H}{ii} continuum.}
}
    \label{tab:two_level}
\end{table}

\begin{figure}
    \resizebox{\hsize}{!}{\includegraphics{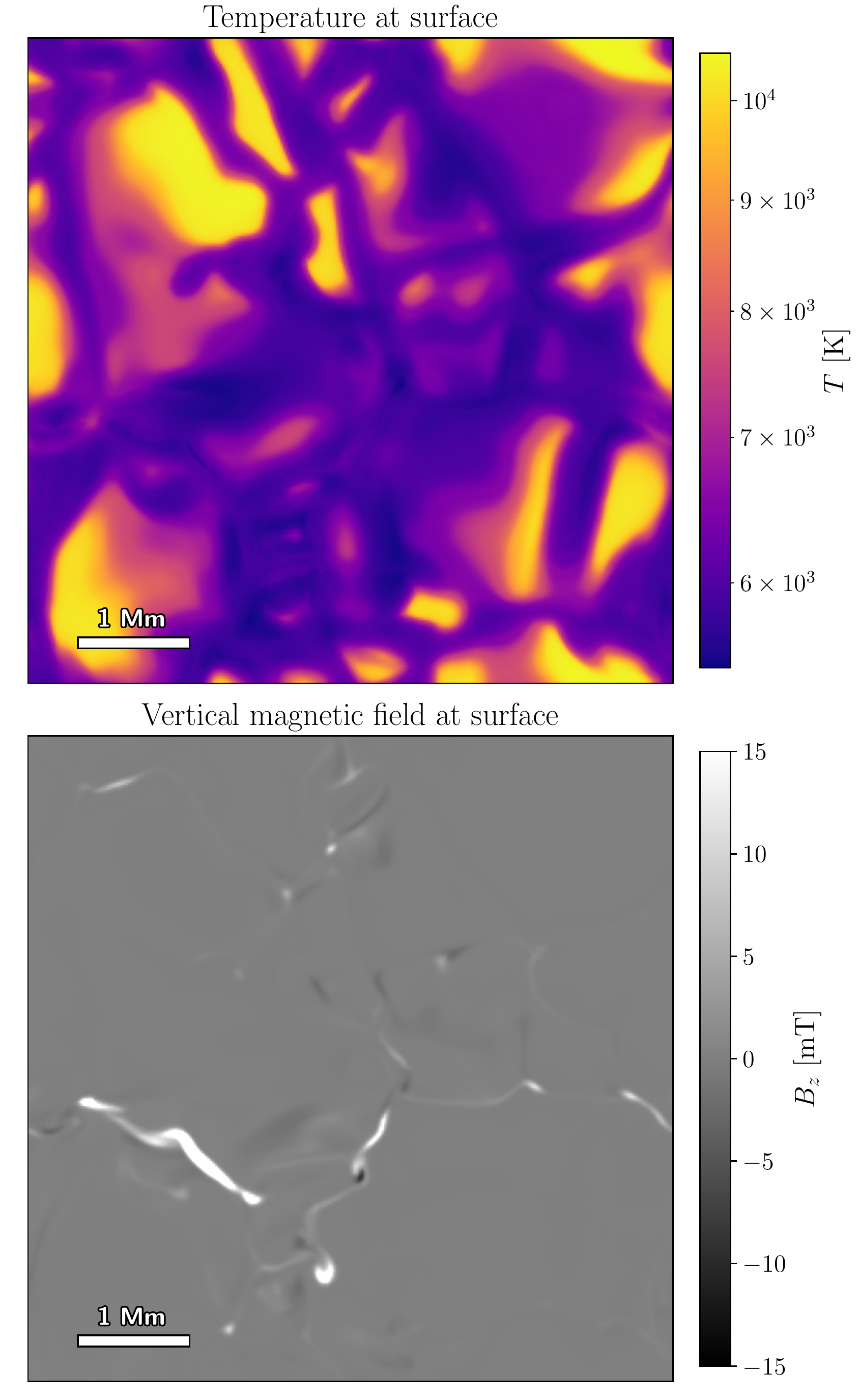}}
    \caption{Surface temperature and vertical magnetic field from the Bifrost simulation used. The horizontal box size is $6\times6$~Mm$^2$.}
    \label{fig:atmos}
\end{figure}

We adopted the level energies and statistical weights shown in Table~\ref{tab:two_level}, along with an $f$-value of 0.41641 and a spontaneous deexcitation rate $A_{ul}$ of $4.7\cdot 10^8$~s$^{-1}$, both taken from the NIST database \citep{NIST:2021}.
The line and continuum extinction were computed using the \texttt{Transparency.jl} package \citep{pereira_transparency}. We calculated the intensity for a total of 90 wavelength points, 50 for the bound-bound transition, and 20 for each of the bound-free transitions. The line profile was computed assuming complete redistribution (CRD) and using standard recipes for van der Waals broadening \citep{Unsold1955}, linear Stark broadening \citep{Sutton:1978wg}, and quadratic Stark broadening \citep{Gray:2005}. For the continuum extinction, we included the following processes: H$^-$ free-free based on \citet{Stilley:1970ws}, H$^-$ bound-free following \citet{Geltman:1962wz}, H free-free as per \citet[][~]{Mihalas:1978book}, H$_2^+$ free-free and bound-free according to \citet{Bates:1952tx}, Rayleigh scattering, and Thomson scattering.

\subsection{Model atmosphere}

As our 3D model atmosphere, we used a single snapshot of a simulation ran with the Bifrost code \citep{Gudiksen:2011vu}. Bifrost solves the magnetohydrodynamic equations on a Cartesian grid, and the simulation employed here included thermal conduction, radiation using a multi-group opacity scheme \citep{Nordlund:1982, Skartlien:2000} including scattering \citep{Hayek:2010}, and a recipe for NLTE radiative losses in the higher layers \citep{Carlsson:2012}. This simulation did not include the effects of hydrogen non-equilibrium ionisation.

The simulation had $256\times 256\times 430$ grid points, or about 28 million cells. These translate into a physical size of $6\times6$~Mm$^2$ in the horizontal direction and 8.7~Mm in the vertical direction. This means that the horizontal spatial pixel size is $23.4\times 23.4$~km$^2$. The vertical grid spacing varied with height, and its average was $20.3\,\unit{km}$. 

The magnetic field configuration of the simulation was chosen to mimic a quiet coronal hole, with a mean unsigned magnetic field at the surface of 0.5~mT and a mostly unipolar configuration. We display the surface temperature and vertical magnetic field in Figure~\ref{fig:atmos}.

\subsection{Constructing the irregular grid}

Our main goal was to investigate whether irregular grids can be used to speed up detailed 3D NLTE radiative transfer calculations. Therefore, a critical step in the process was to construct an irregular grid that is better suited for radiation. There are different ways to build irregular grids, but we employed one of the most general cases: a 3D Voronoi diagram \citep{Voronoi:1908no}.

A Voronoi diagram is calculated from a set of control points called sites. A Voronoi diagram is the set of all cells associated with each site, where a cell is defined as the region that is closer to its control point than to any other control point. The Voronoi cells give us two useful properties for radiative transfer: cell neighbours, and Delaunay lines. Neighbours are sites whose Voronoi cells have a common border, and Delaunay lines are lines connecting neighbouring sites \citep{Delaunay:1934su}. Our ray-tracing algorithm traces rays along Delaunay lines. 

The first step in building a Voronoi diagram from a 3D model atmosphere is to decide where to place the sites, or cell centres, and how many sites to use. As we discuss below, this will have a critical effect on the resulting radiation and speed of convergence. We wish to place a higher density of points in regions that are more important for radiative transfer and therefore follow the distribution of a given quantity.  For a given number of sites, we built the Voronoi diagram using the Monte Carlo technique of rejection sampling. In brief, this involves associating a probability distribution with a reference distribution of the quantity we wish to sample, determining how likely it is to find a control point in a given area. For a given target distribution of sites $f(\vec{r})$, the process is as follows:

\begin{enumerate}
    \item We randomly generated a proposed position $\vec{r}_\mathrm{prop}$ and obtain the proposal density $p_\mathrm{prop} = f(\vec{r}_\mathrm{prop}).$
    \item A random rejection density $p_\mathrm{rej}$ was drawn from a uniform distribution.
    \item The proposal density was compared with the rejection density. If $p_\mathrm{prop} \ge p_\mathrm{rej}$, $\vec{r}_\mathrm{prop}$ was accepted as a control point (site). Otherwise, $\vec{r}_\mathrm{prop}$ was rejected.
    \item We repeated the steps above until the target number of sites was reached.
\end{enumerate}

In our work, the target function $f$ is a quantity that aims to optimise the computational grid. It should have a high value in areas that are important for formation, and lower values in less important areas. We experimented with different quantities from which to sample the location of the sites, but we focus here only on the three most promising ones: the continuum extinction $\alpha^c_\lambda$ at a given wavelength (typically the line core), the log number density of protons under LTE $\log(N_{\ion{H}{ii}}^\mathrm{LTE})$, and the quantity $\log(N_\mathrm{H})^{-2}T^{-2/5}$, a combination of the total hydrogen number density $N_\mathrm{H}$ and temperature.

\begin{figure}
    \resizebox{\hsize}{!}{\includegraphics{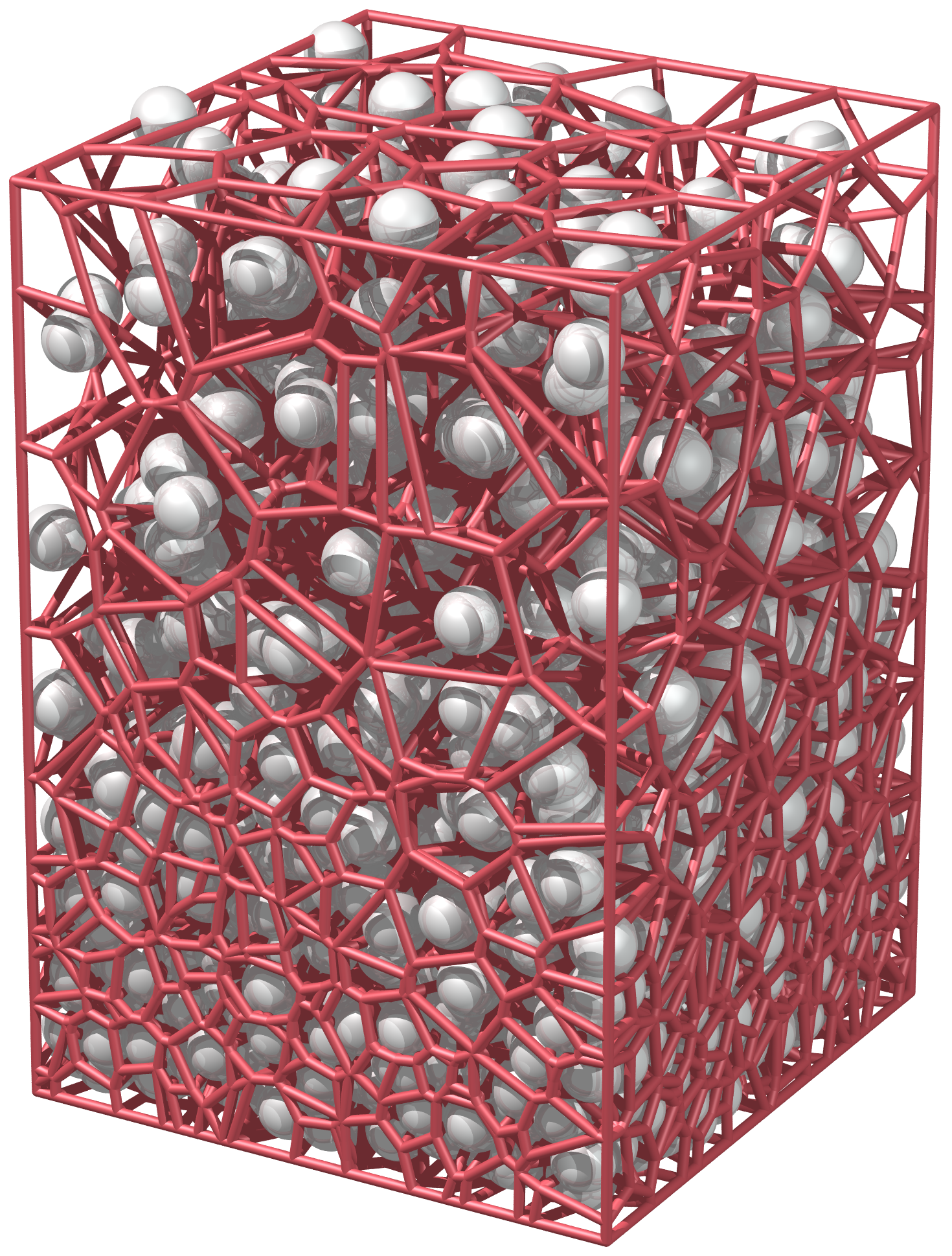}}
    \caption{Illustration of a Voronoi grid sampled from the Bifrost simulation. To clarify some features of the grid, the number of points is significantly lower than the cells in the original grid. This example grid was constructed to follow the log hydrogen density, $\log (N_\textrm{H})$. The white spheres represent the sites, and red lines indicate the edges between cell borders.}
    \label{fig:voronoi_tesselation}
\end{figure}

When the sites of the irregular grid were chosen, we interpolated the atmospheric quantities from the regular grid of the simulation into the new sites using trilinear interpolation. The quantities we interpolated were the temperature, the total hydrogen number density, the electron density, and the velocity vector. 

To construct a Voronoi diagram of the Bifrost simulation, we used the open-source library \texttt{voro++} \citep{Rycroft:2009wq}. This provided the location of the sites and a list of neighbours for all cells, from which we calculated the Delaunay lines that we used for the ray tracing. In Figure~\ref{fig:voronoi_tesselation} we show an example Voronoi grid constructed from our Bifrost simulation, using $\log(N_\mathrm{H})$ to determine the site density. For clarity, this example contains far fewer cells than the original simulation. Still, the site density clearly drops in the higher layers that map the solar corona because its density is much lower. 

\subsection{Rays on irregular grids}

The standard recipes for long characteristics \citep{Jones:1973tm, Jones:1973uu} or short characteristics \citep{Kunasz:1988uj} methods are not directly applicable on irregular grids. To calculate radiation through an irregular grid, we need a method that does not require even partitioning between grid points. Different approaches can be used. \citet{Camps:2013uu} developed a method for tracing rays through straight paths in irregular grids. This has the advantage of  allowing us to trace straight rays through the whole domain for any direction, but requires additional computations to find the walls of each cell and entry and exit points of the ray. \citet{Ritzerveld:2006aa} used a different approach, in which rays were only allowed to travel along Delaunay lines. This approach is also followed in other codes such as \textsc{Lime} \citep{Brinch:2010wj} and \textsc{SimpleX2} \citep{Paardekooper:2010um}. This approach is simpler and requires fewer interpolations, but a downside is that rays can only travel along the predetermined directions that connect cell centres. However, as \citet{Paardekooper:2010um} demonstrated in their Fig. 5, the direction of each segment can be chosen, such as the overall direction (after the ray has travelled through several cells), which remains very close to a target direction. In this work, we followed the approach in which the rays travel only through Delaunay lines because it is computationally simpler.

Our approach to ray tracing was inspired by the short-characteristics (SC) method of \citet{Kunasz:1988uj}. We wished to calculate the intensity at every site, both for upward-travelling rays starting at the bottom boundary and for downward-travelling rays starting at the top boundary. While \textsc{Lime} and \mbox{\textsc{SimpleX2}} use a Monte Carlo approach to sample different directions, we instead adopted a fixed-angle quadrature, which we used to integrate the intensity in different directions and obtain $J_\lambda$. \citet{Stepan:2020wf} and \citet{Jaume-Bestard:2021vi} have developed optimised angle quadratures for 3D radiative transfer in stellar atmospheres, and here we adopted the quadrature optimised for unpolarised radiative transfer with $n_\mathrm{rays}=12$ and $L=7$ from \citet{Jaume-Bestard:2021vi}.

\begin{figure}
    \includegraphics[width=0.75\linewidth]{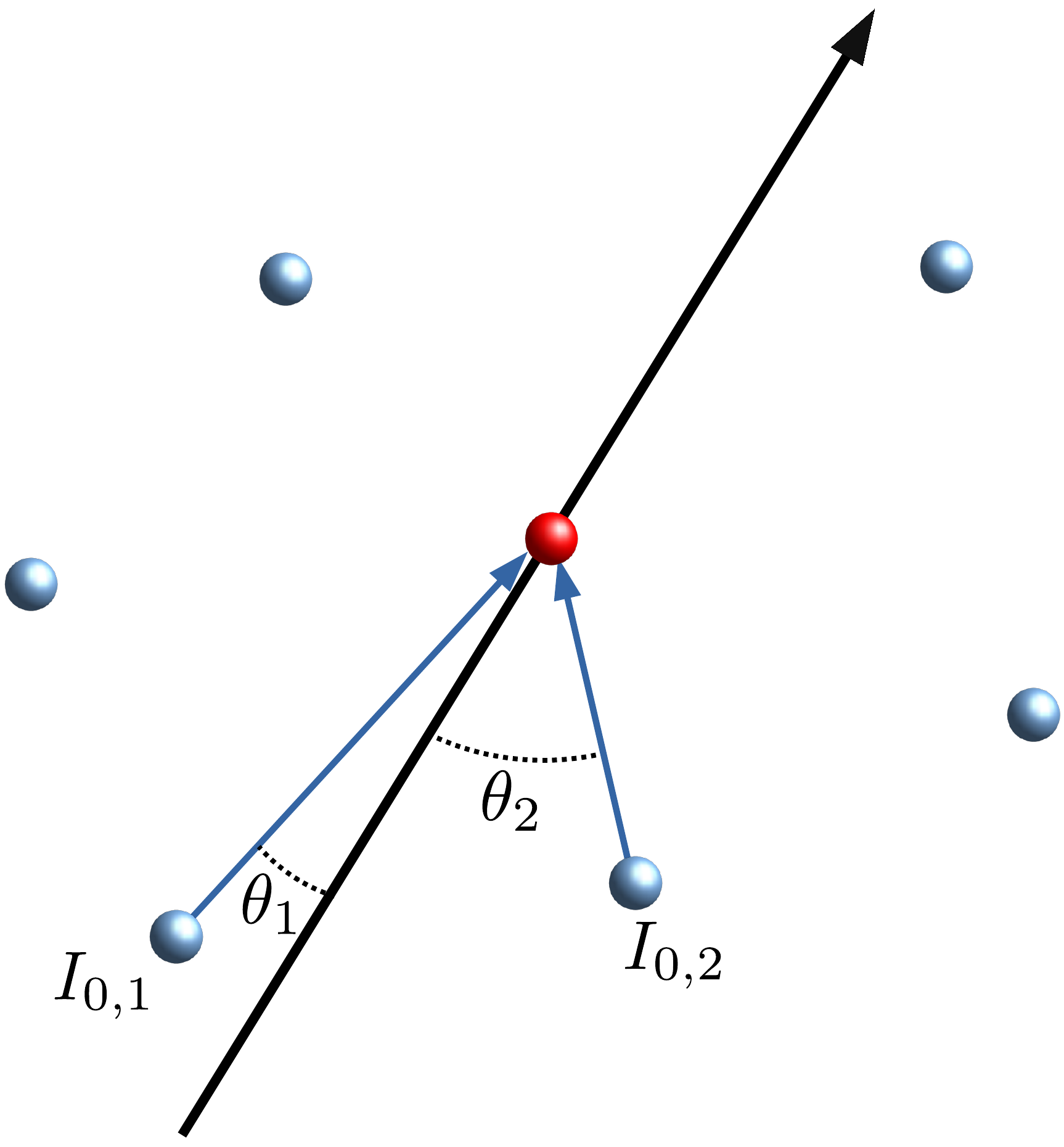}
    \caption[Two-dimensional representation of ray tracing in an irregular grid]{Two-dimensional representation of ray tracing in an irregular grid. We wish to calculate the intensity at the red point, and the thick black arrow is the direction of the ray (characteristic). In our algorithm, the incoming intensity is a combination of $I_{0,1}$ and $I_{0,2}$, the intensity coming from the two sites with the smallest angles between the Delaunay lines and the direction of the ray, $\theta_1$ and $\theta_2$.}
    \label{fig:delaunay}
\end{figure}

For each ray in the quadrature, we computed the intensity by tracing rays along the Delaunay lines that lie closest to the characteristic angle. In some parts of our atmosphere, the irregular grid can have sparse sampling, with large cells. In these cases, limiting the transport to a single Delaunay line can lead to errors in the formal solution. To better sample the grid, we implemented a method for approximating the intensity by using not only a single Delaunay line, but the two Delaunay lines that have the smallest angle from the characteristic ray,
\begin{equation}
    I_i = w_1 I_{i,1} + w_2 I_{i,2}\,,
    \label{eq:delaunay_rays}
\end{equation}
where we designed the weights $w_i$ to favour Delaunay lines that have a smaller angle to the target quadrature angle, and normalised them,
\begin{align}
    w_i = \frac{v_i^{\scalebox{1.025}{$r$}}}{\sum_{j=k_1,k_2}v_j^{\scalebox{1.025}{$r$}}} \,,
\end{align}
where $v_i$ is the dot product between the direction of the ray and the direction of the Delaunay line, and the exponent $r$ is treated as a free parameter. After some experimentation, we find that $r=7$ gives reasonable results, but overall, the method is not very sensitive to the particular choice of $r$.

The transport of radiation can be designed in different ways (e.g. combining the intensity from additional neighbours or adopting different weights), but we find that our algorithm works well to correct the  discrepancy between a quadrature angle and the angles of the Delaunay lines. In an irregular grid with sites drawn from a 3D Poisson distribution, a site has 15.54 neighbours on average \citep{van-de-Weygaert:1994ua}. With our 3D atmosphere with $\approx300\,000$ sites drawn according to the hydrogen density, the average number of neighbours was 15.44. Consequently, every site should have some Delaunay lines whose direction differs from a quadrature ray by only a small amount.

\begin{figure*}
    \sidecaption
    \includegraphics[width=12cm]{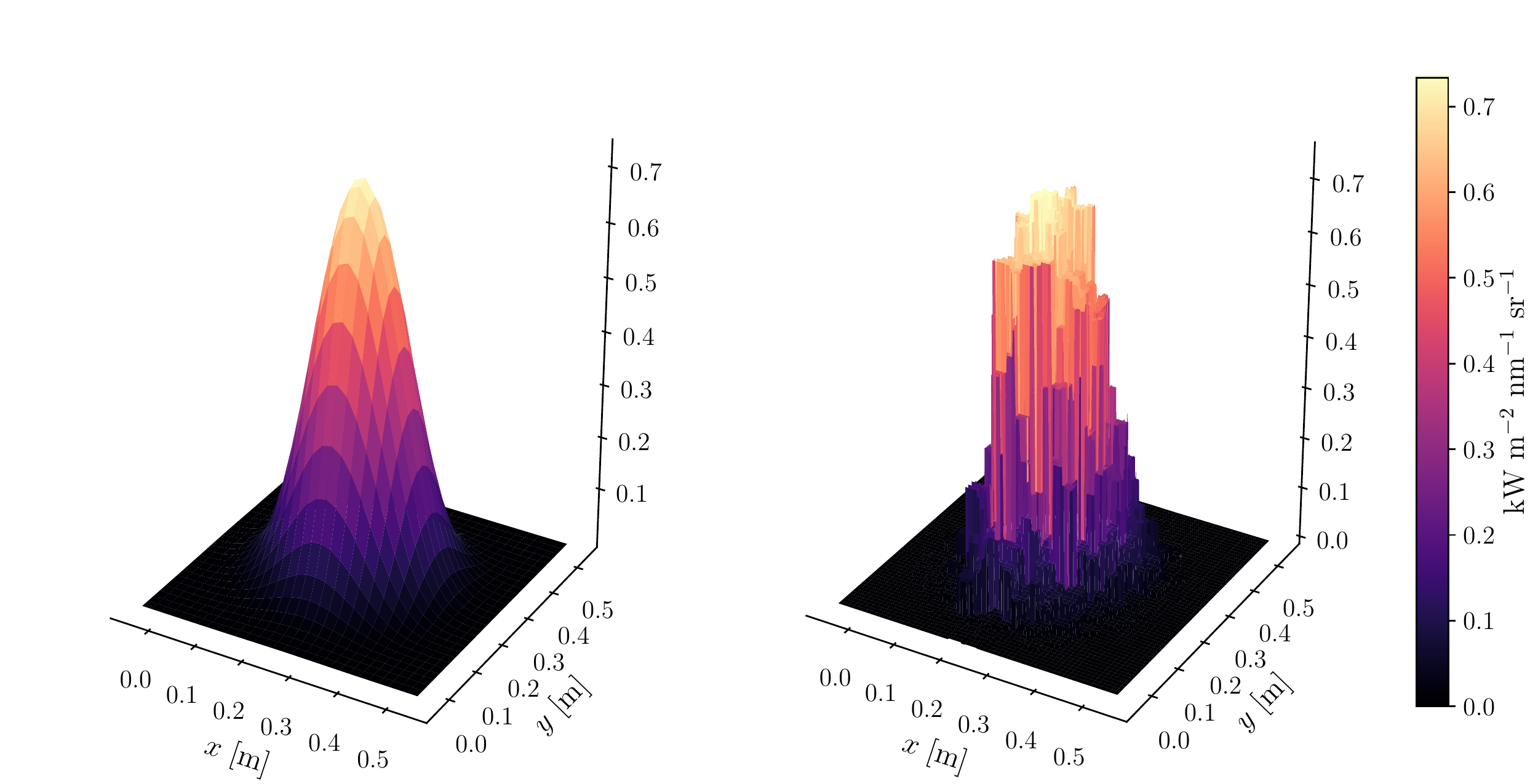}
    \caption{Emerging intensities from searchlight beam tests. \textit{Left:} Regular grid with short characteristics and linear interpolation. \textit{Right:} Irregular grid. The box we used had a size of $1\times 1\times 1~\unit{m}^3$ and $51^3$ grid points. The beam was circular, with a radius of $0.2~\unit{m}$ and inclination of $\theta = 20\unit{\degree}$ ($\mu\approx0.94$).}
    \label{fig:searchlight_3D}
\end{figure*}

To compute $I_\lambda$ along each Delaunay line, we solved the integral in the formal solution (\ref{eq:Formal}) by using linear interpolation of $S_\lambda$. This accuracy can be improved by using higher-order schemes \citep[e.g.][]{de-la-Cruz-Rodriguez:2013aa, Janett:2019ab}. However, given the exploratory nature of this work and several of the approximations already used, the additional accuracy of higher-order methods appeared to be not worth the additional computational complexity.

Our algorithm for computing $I_\lambda$ along different rays was used to obtain $J_\lambda$, which we then used in the $\Lambda$-iteration to obtain the source function and the level populations. After the iteration converged, the end product should be the emergent $I_\lambda$ at a given direction. To obtain $I_\lambda$ , we could have used the same ray-tracing algorithm with the converged level populations. However, this would create undesirable effects in this type of simulation. 
A side-effect of placing the number of sites in the locations that are most important for the transport of (optically thick) radiation is that the upper layers of the simulation box, covering the corona, have far fewer cells. This can be seen, to some extent, in Fig.~\ref{fig:voronoi_tesselation}. This means that in the irregular grid, $I_\lambda$ at the top of the simulation box, comes from a small number of sites. This results in poor spatial resolution. Therefore, instead of using the irregular grid to compute the final $I_\lambda$, we instead interpolated the converged level populations to the original grid of the simulation and then calculated a formal solution using long characteristics. The choice of interpolation mapping the irregular grid back to the original mesh is important for the intensity calculations. We used inverse distance interpolation with the two nearest neighbours because it reduces noise from the random sampling of the grid and preserves sharp features in intensity. The interpolation allowed us to obtain $I_\lambda$ at the original grid spacing of the simulation. For the tests performed here, we compared only vertical rays ($\mu=1$), but any viewing angle can be used with this method.

\subsection{Implementation} 

Our implementation of the methods described above is open source and publicly available \citep{Udnaes:2022co}\footnote{The latest version is available at \url{https://github.com/meudnaes/VoronoiRT}.} Our code is written in the programming language Julia \citep{Julia:2017}. Its relatively simple architecture makes use of Julia's thread parallelism, which can run across different cores in a shared memory architecture. The code is not domain decomposed and does not make use of more advanced parallel architectures.

For a fair comparison with existing radiative transfer methods, our code also included an implementation of the short-characteristics scheme of \citet{Kunasz:1988uj} that we ran on a regular grid. This allowed us to run exactly the same functions to calculate extinction, collisional and radiative rates, $\Lambda$-iteration, statistical equilibrium, and the final formal solution. The only difference is the method for computing $J_\lambda$, which can be selected as regular grid or irregular grid. Hereafter, when we refer to calculations on the ``regular grid'', we mean calculations performed using short characteristics in a regular grid, using linear interpolation to obtain the simulation quantities at the necessary locations.

\section{Results\label{sec:results}}

\subsection{Searchlight beam test}

Our first test of our ray tracing in irregular grids was a searchlight beam test. In this test, we created a box in which the extinction $\alpha_\lambda$ and source function $S_\lambda$ were zero everywhere and propagated a radiation beam from the bottom to the top of the box at a given angle. We chose a box size of $1\times 1\times 1~\unit{m}^3$ and a total of $51^3$ sites. From the centre of the bottom boundary, we propagated a circular beam with $I_\textrm{beam} = 1~\iunits$, and a radius of $0.2~\unit{m}$. 

For this test only, the Voronoi cells were randomly sampled, and not sampled from a simulation quantity. If we had sampled the sites from a quantity such as mass density, the density of sites at the top of the box (where we assess the results) would be too low, and it would be a poor test of the diffusion introduced by the algorithm. Therefore, the $51^3$ sites were sampled from a uniform random distribution to ensure a similar density throughout the box.

In Fig.~\ref{fig:searchlight_3D} we show the results of a beam with an inclination of $20\unit{\degree}$ from the vertical ($\mu\approx0.94$). We compare them with the results from the regular grid. Ray tracing on the irregular grid compares very favourably with results from the regular grid. The amount of diffusion introduced by the irregular grid solver is comparable with that of short characteristics with linear interpolation. For this setup, the peak intensity of the beam is reduced to about 70\% of the original value in both cases. For shallower rays, the diffusion in regular grids increases \citep{de-Vicente:2021wk}, and again we find comparable results for the irregular grid.

\subsection{Radiative transfer on irregular grids in LTE}
\begin{figure}
    \sidecaption
    \resizebox{\hsize}{!}{\includegraphics{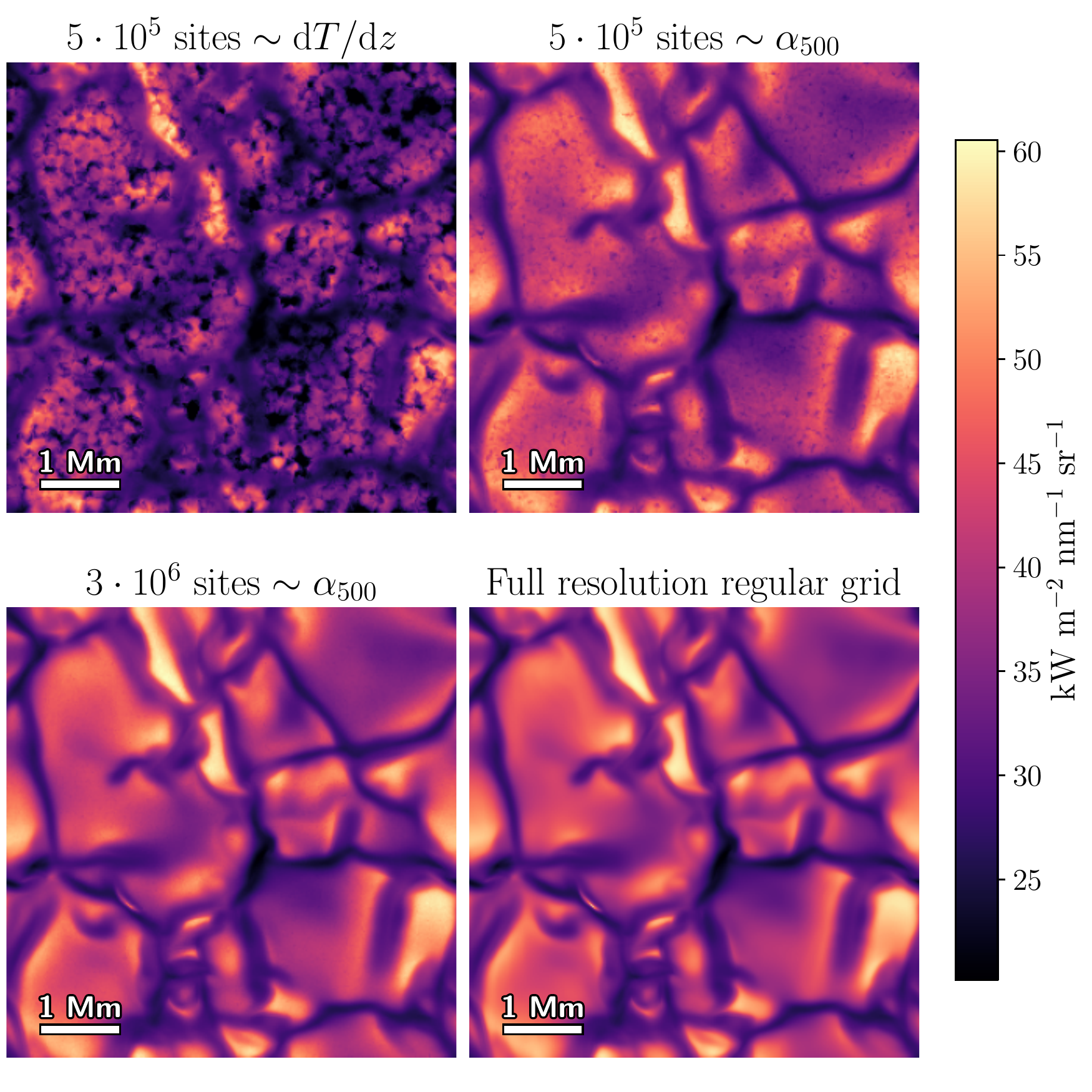}}
    \caption{Disk-centre continuum intensity at 500~\unit{nm} from the Bifrost simulation. Each panel was computed using a different grid. The bottom right panel shows the regular grid, and the other three panels show irregular grids with a different number of sites and constructed sampling distributions of $\mathrm{d}T/\mathrm{d}z$ and $\alpha_{500}$, the continuum extinction at 500~nm. All calculations assumed LTE.}
    \label{fig:LTEmosaic}
\end{figure}

The next step was a calculation of intensity in LTE. Strictly speaking, this was no test of the irregular grid solver because no radiation was calculated with the irregular solver. Instead, it was a test of the procedure to construct an irregular grid and to interpolate quantities from the regular to irregular grid and vice versa.

A NLTE problem revolves around iterating the source function until the radiation field is consistent with the atomic level populations. The populations from a converged $\Lambda$-iteration are the end product that is then used to carry out spectral synthesis. In this LTE test, our procedure was to start by computing the LTE populations for our hydrogen-like atom in LTE in the original grid of the Bifrost simulation. Next, we built an irregular grid and interpolated the populations into to the irregular grid. As no iteration took place, the populations were then interpolated back into the original regular grid. Finally, we computed the emerging continuum intensity at 500~nm for $\mu=1$ from the regular grid using the interpolated populations. The source function was computed in LTE using only quantities in the original grid, but the extinction and optical depth make use of the interpolated populations because we used them as hydrogen populations, and therefore, they were key in setting the H$^{-}$ populations, whose bound-free and free-free contributions dominate the continuum extinction.

For this test, we built the irregular grids using two methods: by sampling $\alpha_{500}$, the continuum extinction at 500~nm, and by sampling the vertical temperature gradient. We built grids with a different number of sites, from $5\times 10^5$ to $3\times 10^6$ sites. For comparison, the original grid has $2.8\times 10^7$ grid points.

We show the results of the continuum intensity in LTE in Fig.~\ref{fig:LTEmosaic}. For this case, it is obvious that sampling the irregular grid from $\alpha_{500}$ produces much better results than sampling from the vertical gradient of temperature because the noise from the interpolation is much less noticeable. With the grid sampled by $\alpha_{500}$,  the granulation pattern, overall contrast, and absolute intensities become close to the result of the regular grid already with only $5\times 10^5$
sites (almost two orders of magnitude fewer grid points than the original
grid). With $3\times 10^6$ sites, the results from the irregular grid are virtually indistinguishable from the regular grid. This is encouraging as it shows that the irregular grid can sample the continuum-forming region very well, allowing calculations with nearly ten times fewer grid points.

\subsection{Radiative transfer on irregular grids in NLTE}

Our final and most extensive step was to do a full NLTE calculation using our simplified model atom with one spectral line and two bound-free edges. With the collisional rates artificially scaled up, we reduced the amount of scattering and ensured that the problem was solvable by using the simple $\Lambda$-iteration. Since our code for these experiments has only simple parallelism, running the full 3D NLTE problem is still computationally expensive.

\begin{figure}
    \resizebox{\hsize}{!}{\includegraphics{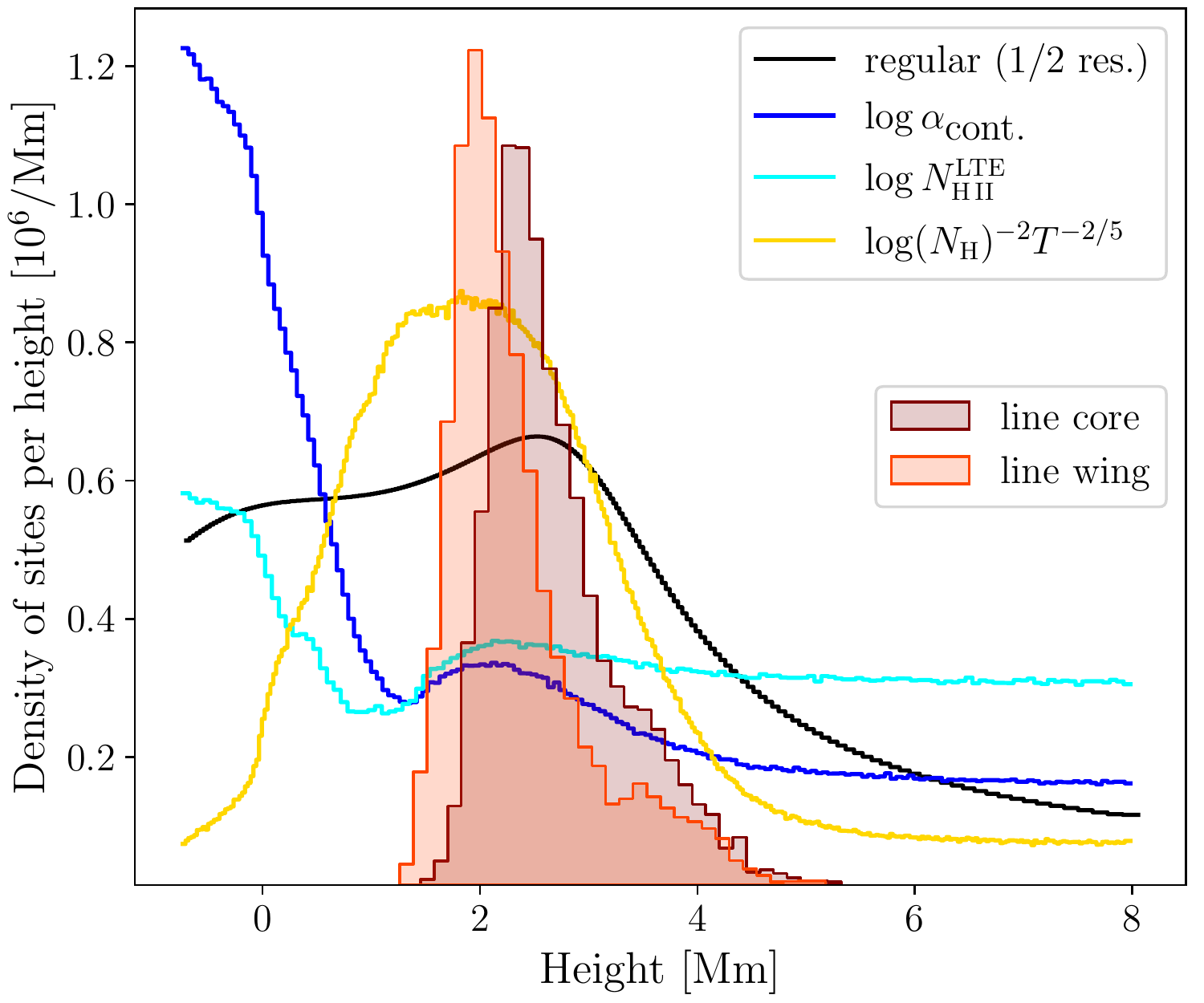}}
    \caption{Distributions of the vertical density of the grid points for different grids. The regular grid (\textit{black}) was taken at half resolution, meaning every second grid point. The different irregular grids (\textit{blue}, \textit{cyan}, and \textit{yellow}) were sampled by the quantity indicated in the legend and had three million sites. The shaded histograms (\textit{red}) show the $\tau=1$ heights at the line core and the wing at $-27~\unit{km.s^{-1}}$ from the core.}
    \label{fig:grid_vs_formation}
\end{figure}

\begin{figure*}
    \includegraphics[width=17cm]{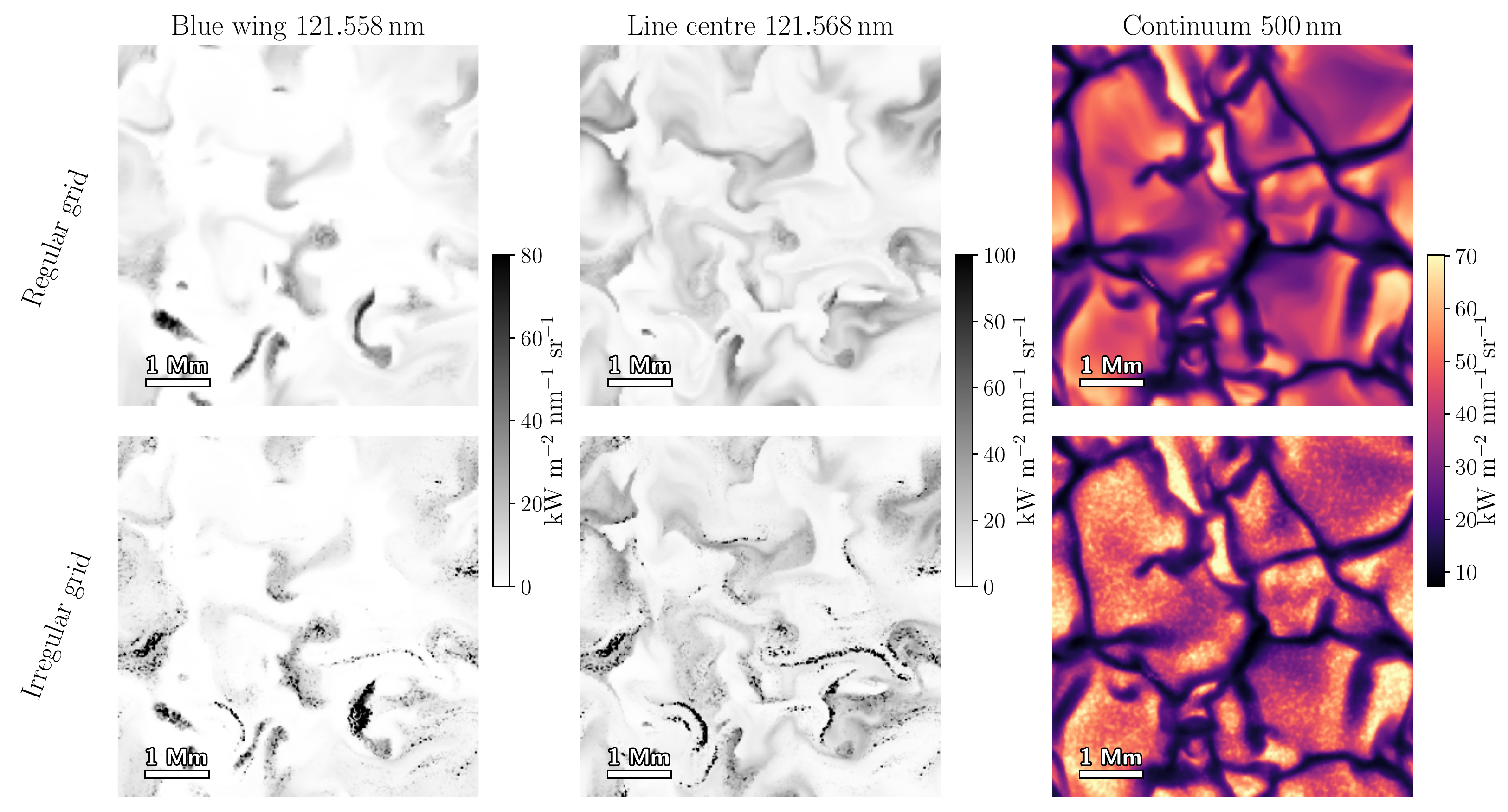}
    \caption[Disk-centre intensity for three different wavelengths from regular and irregular grids]{Disk-centre intensity for three different wavelengths calculated with the half-resolution regular grid (\textit{top panels}) and with an irregular grid with three~million sites sampled from $\log(N_\textrm{H})^{-2}T^{-2/5}$ (\textit{bottom panels}). The blue wing position is at $-27\,\unit{km.s^{-1}}$ from the line centre.}
    \label{fig:disk_centre}
\end{figure*}

To avoid lengthy running times, we did not run the experiment at the native resolution of the Bifrost simulation (or equivalent number of sites in the irregular case) because this greatly increased not only the time per iteration, but also the total number of iterations necessary to achieve convergence. Instead, we ran the regular grid at one-half resolution, which we defined as taking every second point from the original grid in all three directions. This means that the half-resolution grid has eight times fewer grid points, or about 3.5 million points. For the corresponding irregular grid test, we used three million sites. In addition, we also ran tests using  one-third and one-fourth resolution, and the same number of sites in the irregular cases, but their main purpose was to time the differences.

We experimented with different irregular grids that sampled the distribution of three quantities: $\log_{10}(\alpha_\mathrm{cont})$, the logarithm of the continuum extinction at the line core; $\log(N_\ion{H}{ii}^\mathrm{LTE})$, the logarithm of the proton density under LTE; and $\log(N_\mathrm{H})^{-2}T^{-2/5}$, where $N_\mathrm{H}$ is the total hydrogen density, and $T$ is the temperature.
The rationale for these three choices was as follows. The continuum extinction at the line core is derived mostly from quantities computed in LTE such as the H$^{-}$ populations and is proportional to the mass density. Following the results from the LTE radiative transfer tests, this results in a grid that is optimised for the line continuum wavelengths (but not necessarily for the line profile). The proton density under LTE follows the mass density, but also the electron density and temperature via the Saha-Boltzmann distribution, and it is more sensitive to changes in the upper layers than $\alpha_\mathrm{cont}$. Finally, $\log(N_\mathrm{H})^{-2}T^{-2/5}$ was an ad hoc formulation found by a little experimentation, with the goal of placing a denser grid in the regions most critical to the line formation. Ideally, an irregular grid would be built by sampling the quantities that are most relevant for radiation at all the wavelengths considered, for example the line extinction or level populations. However, this approach is moot because the values of these quantities are not known before the NLTE calculations, so that we are restricted to the quantities that are available at the start of the calculations (or that are derived from these quantities). 

In Fig.~\ref{fig:grid_vs_formation} we compare the average density of sites or grid points per height as a function of height in the simulation. We also plot histograms of the height of the optical depth unity for the line core and line wing, which is a proxy for the formation height at these wavelengths. The regular grid has a variable height resolution and therefore has a higher resolution in the photosphere and chromosphere than in the corona. The irregular grid sampling $\log_{10}(\alpha_\mathrm{cont})$ has a higher density of sites in the photosphere, while $\log(N_\ion{H}{ii}^\mathrm{LTE})$ follows a similar curve, but has a lower resolution in the deeper layers and a better resolution from about 2~Mm to the top of the box. The grid following $\log(N_\mathrm{H})^{-2}T^{-2/5}$ has the lowest resolution of all in the photosphere, but a larger number of sites between 1--3~Mm, covering the region of formation of our modelled spectral line. 

\begin{figure}
    \resizebox{\hsize}{!}{\includegraphics{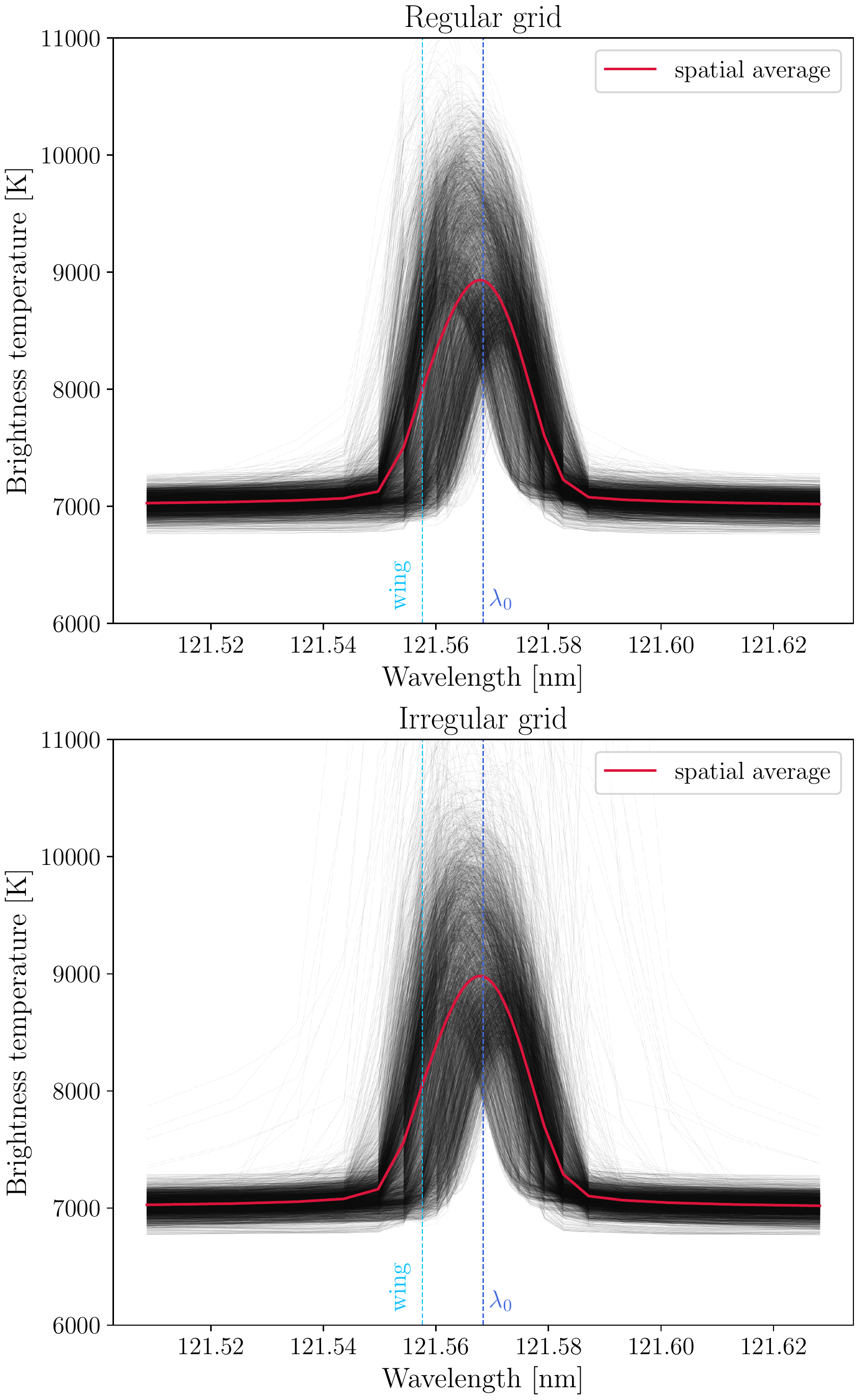}}

    \caption{Individual line profiles for the disk-centre intensity calculated from the half-resolution regular grid and an irregular grid with three~million sites sampled from $\log(N_\textrm{H})^{-2}T^{-2/5}$. Every black line corresponds to the brightness temperature from a column in the atmosphere. The spatial average over all columns is represented by the red line. The dashed blue lines indicate the wavelength positions of the blue wing at 121.558~nm (or $-27~\unit{km.s^{-1}}$) and line centre $\lambda_0$ at $121.568\,\unit{nm}$.}
    \label{fig:lines}
\end{figure}

We find that the irregular grid sampled by $\log(N_\mathrm{H})^{-2}T^{-2/5}$ gives the most accurate results when compared with the regular grid. In Fig.~\ref{fig:disk_centre} we show emergent intensity images at $\mu=1$ for both the regular and irregular grid at three wavelengths: the line wing, the line core, and the 500~nm continuum, and in Fig.~\ref{fig:lines} we show the $\mu=1$ spectral profiles. For the sake of brevity, we omitted the results from irregular grids sampled by $\log_{10}(\alpha_\mathrm{cont})$ and $\log(N_\ion{H}{ii}^\mathrm{LTE})$. 

\begin{table}
    \centering
    \caption{Time per iteration for irregular and regular grids at different resolutions.}
    \begin{tabular}{r r r c} 
    \hline\hline
    Grid points & Regular grid & Irregular grid & ratio\\
                &    (s)           &     (s)            & \\[0.5ex] 
    \hline
    $442\,368$ & $600$ & $2\,240$ & 3.73\\ 
    $1\,065\,024$ & $1\,437$ & $5\,372$ & 3.74\\
    $3\,522\,560$ & $4\,621$ & $18\,217$ & 3.94\\
    \hline
    \end{tabular}

    \label{tab:time}
\end{table}

Fig.~\ref{fig:disk_centre} shows that the results are very similar overall in shape and intensity level, but that the irregular grid shows some spurious results of points with very high intensities in the line wing and core (black dots). These points are more visible in the line core images, but are still a small subset of the total number of points. They are also visible in Fig.~\ref{fig:lines} as the very broad and strong profiles. Nevertheless, most profiles agree very well between the irregular and regular grid calculations. The occurrence of the spurious strong profiles seems to be related to how well the irregular grid covers the line-forming region. They are much more numerous when we use fewer sites (e.g. $10^5$ instead of $3\times 10^6$), and they are also more numerous when we use irregular grids sampled from $\log_{10}(\alpha_\mathrm{cont})$ or $\log(N_\ion{H}{ii}^\mathrm{LTE})$ (not shown).

The 500~nm continuum images of Fig.~\ref{fig:disk_centre} also show that the irregular grid produces a noisy map. This is because the irregular grid sampling $\log(N_\mathrm{H})^{-2}T^{-2/5}$ has far fewer sites in the photosphere, and the noise recedes with the grids built by sampling $\log_{10}(\alpha_\mathrm{cont})$ or $\log(N_\ion{H}{ii}^\mathrm{LTE})$.

\begin{figure}
    \resizebox{\hsize}{!}{\includegraphics{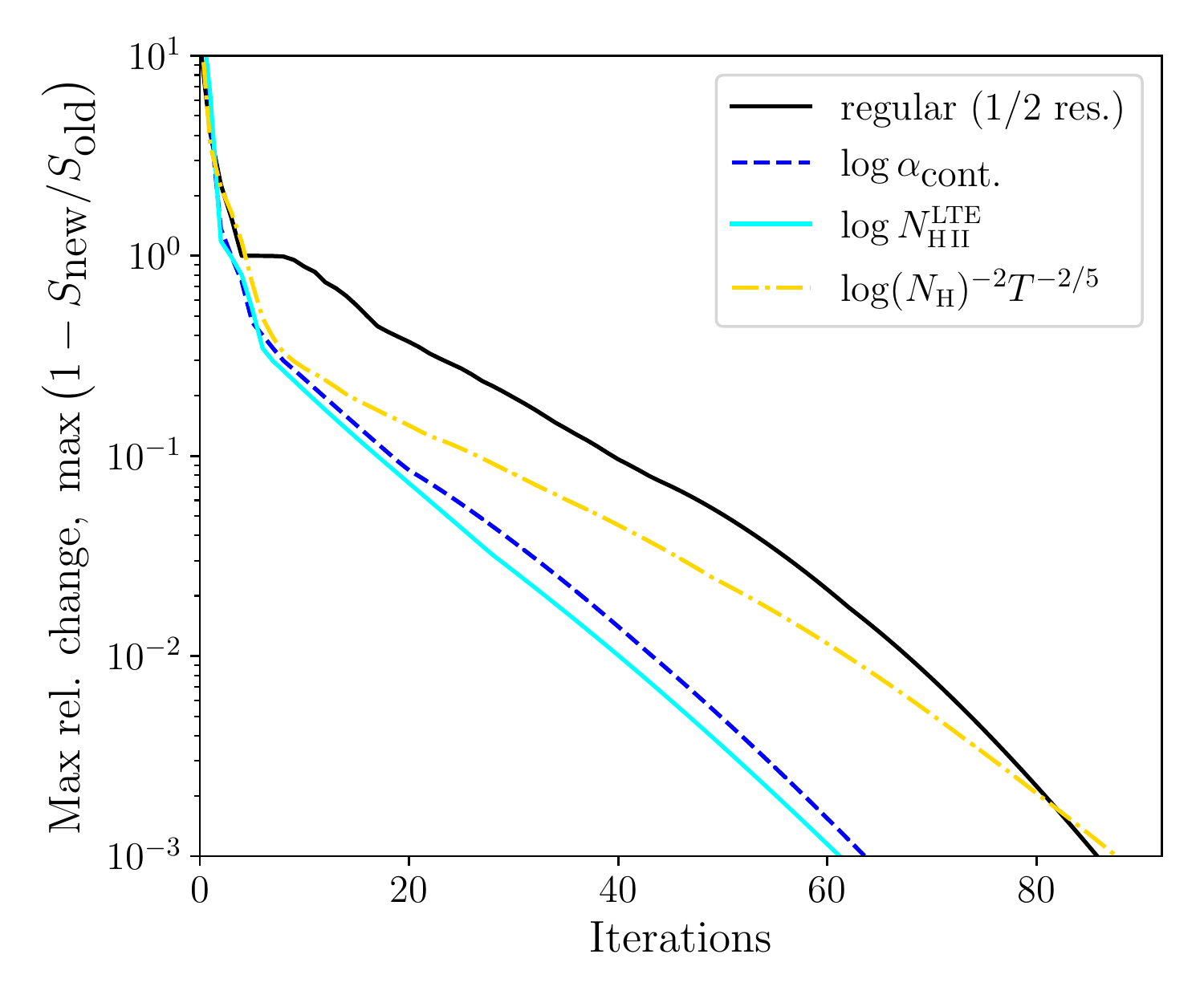}}
    \caption{$\Lambda$-iteration convergence for irregular grids with three~million sites constructed from the labelled quantities, and the half-resolution regular grid of the Bifrost atmospheric model. }
    \label{fig:convergence}
\end{figure}

In addition to the accuracy of the results, an important aspect of this analysis is to determine whether irregular grids can save computer time when compared to calculations in regular grids. For the computational costs of running a single iteration, we list in Table~\ref{tab:time} the time it took to run a single iteration on the regular and irregular grids for different numbers of grid points. The regular grid experiments were performed at one-half, one-third, and one-quarter resolution, and for each case the corresponding irregular grids had exactly the same number of grid points (sites). We performed the experiment using five cores of an AMD EPYC 7763 2.45GHz CPU and averaged the times over three iterations. There was no difference in time per iteration from the different irregular grids. 

We plot in Fig.~\ref{fig:convergence} the convergence rate versus the number of iterations for the three irregular grids and the regular grid to show how many iterations were needed to achieve convergence. These 3D NLTE problems are usually assumed to converge at a level around $10^{-3}$. The grid sampling $\log(N_\mathrm{H})^{-2}T^{-2/5}$ converges in about the same number of iterations as the regular grid, while the other two converge with about one-third fewer iterations. 

\section{Discussion}

Our main aim with this work was to determine whether it is feasible to use irregular grids to speed up 3D NLTE radiative transfer, just like optimised height grids can be used to speed up 1D NLTE calculations. We employed one of the most general irregular grids, based on a 3D Voronoi diagram. In this first exploration, we employed a simplified approach with an algorithm that is not massively parallel and runs only on shared-memory systems. To further reduce the complexity of the code, we also used the simpler $\Lambda$-iteration scheme, which breaks down in the strong scattering regimes. To reduce computation times, we used a simplified hydrogen-like model with two bound states and one continuum and assumed complete redistribution. To make the problem tractable given these simplifications, we artificially increased the collisional rates. 

We focus on the spectral profiles of the line from our model atom, a modified Lyman-$\alpha$ line. Because we reduced the amount of scattering by increasing the collisional rates, the source function is closer to LTE. Compared to Lyman-$\alpha$, our line is formed in hotter regions and lacks the central reversal. This explains why in both absolute intensity units and in brightness temperature it is consistently brighter than Lyman-$\alpha$ from observations or simulations \citep[e.g.][]{Schmit:2017}.

Given the assumptions above, our experiments still give us valuable insight. Our searchlight beam test showed that our ray-tracing algorithm in irregular grids produces good results, showing a diffusion that is comparable to short characteristics with linear interpolation. The test of synthesising continuum intensities in LTE probed the effects of interpolating to a Voronoi grid and back to the original grid, and it gave two key results. First, it demonstrated that it is crucial to build the irregular grid on a quantity relevant for the transport of radiation, with $\alpha_{500}$ being the most relevant quantity here. Second, it showed that a properly optimised irregular grid can facilitate the calculation of continuum radiation with about an order of magnitude fewer grid points than the original grid. However, we recall that continuum radiation comes from a relatively shallow region, so that similar gains can probably be achieved with regular grids by limiting the $z$ range. A much more stringent test is the NLTE formation of our spectral line, where our results are more mixed.

When we solved the 3D NLTE problem for our model atom, we had to compute radiation for 90 wavelengths, covering both the spectral line and bound-free edges, and solve the statistical equilibrium equations. This test is more stringent on the irregular grid because the range of formation heights is much broader, so that an ideal irregular grid should provide good resolution for wavelengths from the line core to the continuum, which span a formation region of more than 4~Mm. Finding a relevant quantity from which to sample the irregular grid in this case is a much more challenging task. Especially because many of the relevant quantities (e.g. line extinction) are not known a priori, before we solved the NLTE problem. The design of the irregular grid has dramatic effects on the resulting populations and line profiles in this case as well.

Irregular grids that sample the distribution of continuum extinction or LTE proton density lead to unsatisfactory results: The emergent line profiles show many instances with excessive intensities that show up noticeably in intensity maps. This is perhaps to be expected from Fig.~\ref{fig:grid_vs_formation}: The grid resolution in these cases is much lower than the regular grid in the regions in which the line is formed. It appears that the spurious line profiles arise from large jumps in the populations when interpolating from coarser grids. To address this issue, we built an irregular grid sampling the seemingly arbitrary quantity $\log(N_\mathrm{H})^{-2}T^{-2/5}$, which led to a much higher density of sites in the line-forming region. With this irregular grid, we still obtained a few spurious profiles (as seen in Figs.~\ref{fig:disk_centre}-\ref{fig:lines}), but they are far less prevalent, and the profiles and intensity level agree well with the results from the regular grid overall. A side-effect from this particular grid is that the photospheric radiation is adversely affected by noise from the interpolation because this grid has a lower resolution in the photosphere. Despite these issues, our results show that it is possible to use irregular grids for 3D NLTE radiative transfer. The next important question is therefore whether this leads to faster run times than regular grids.

We find that one iteration on a irregular grid is about four times slower than one iteration in the regular grid. This is comparable with the results of \citet{Camps:2013uu}, who reported that radiative transfer in Voronoi grids was about three times slower than on octree grids (and radiation transport across octree grids is already more complex than our regular grid). However, we find that our most reliable irregular grid reached convergence in about the same number of iterations as the regular grid ($\approx 90$ iterations to achieve a $10^{-3}$ level). This means that in this case, irregular grids are much slower to solve the 3D NLTE problem. Some irregular grids can converge faster, in about two-thirds of the regular grid iterations, but the lower number of iterations is not enough to balance the price penalty of iterations that are four times slower. Tellingly, the irregular grids that converge faster have a lower density of grid points in the line-forming region, which leads us to conclude that they converge faster precisely because there are fewer grid points through which the changes to the source function need to be propagated. These faster-converging grids also have the more serious problem of leading to additional spurious line profiles. 

\section{Conclusions}

We conclude that irregular grids alone do not provide the optimal solution for efficient 3D NLTE calculations. With our setup, they are about four times slower and lead to less accurate results. The design of an irregular grid that lends itself to radiative transfer across a large height range is not trivial and affects both the reliability of the results and the speed of convergence. It may be possible to build a better-performing irregular grid sampling other quantities, but this work goes beyond our exploratory analysis. NLTE radiative transfer in irregular grids is still affected by the curse of dimensionality, just like regular grids, where going to 3D and increasing the spatial resolution leads to slower convergence. 

Irregular grids might still be attractive through their potential to achieve results similar to a regular grid with fewer grid points. Based on our results for the LTE continuum intensity, we speculate that that the NLTE results from our irregular grid with $3\times 10^6$ sites, when interpolated back to the original simulation grid, could give comparable results to a calculation on the full resolution regular grid with $\approx 3\times 10^7$ grid points. We did not test this due to the long running times of the full resolution in a regular grid. These regular grid full-resolution runs should be about eight times slower than our half-resolution runs, and in these cases, it is possible that the irregular grid running times would outweigh the four times slower iterations.

\begin{acknowledgements}
We would like to thank Mats Carlsson for helpful suggestions, fruitful discussions, and for providing the Bifrost atmosphere used in this paper. We would also like to thank Jaime de la Cruz Rodriguez for helpful discussions and suggestions for improvements. This work has been supported by the Research Council of Norway through its Centers of Excellence scheme, project number 262622. We kindly acknowledge the computational resources provided by UNINETT Sigma2 - the National Infrastructure for High Performance Computing and Data Storage in Norway.
\end{acknowledgements}

\bibliographystyle{aa}
\bibliography{references}

\end{document}